# Latest results on light (anti)nuclei production in Pb–Pb collisions with ALICE at the LHC

**Chiara Pinto**[a,b,*] **on behalf of the ALICE Collaboration**

[a] *Department of Physics and Astronomy "E. Majorana", University of Catania,*
 *Via S. Sofia 64, Catania, Italy*

[b] *INFN, Section of Catania,*
 *Via S. Sofia 64, Catania, Italy*

*E-mail:* chiara.pinto@cern.ch

Recent results on the production of light nuclei, including deuterons, tritons, $^3$He, $^4$He and the corresponding antinuclei in Pb–Pb collisions at $\sqrt{s_{NN}}$ = 5.02 TeV are presented and compared with theoretical predictions to provide insight into their production mechanisms in heavy-ion collisions. The large variety of measurements performed with the ALICE apparatus at different energies and collision systems allows us to constrain the models of the production mechanisms of light flavour baryon clusters, in particular those based on the coalescence and statistical hadronisation approaches.

Furthermore, new measurements of the elliptic and triangular flow of deuteron and $^3$He produced in Pb–Pb collisions at $\sqrt{s_{NN}}$ = 5.02 TeV are presented and compared to the expectations from coalescence and hydrodynamic models. The measurement of the elliptic and triangular flow of light nuclei provides a powerful tool to give insight into their production mechanism and freeze-out properties at a late stage of the collision evolution.

*Speaker

## 1. Physics motivation

The production mechanism of light (anti)nuclei in heavy-ion collisions is a subject of intense debate. The binding energy of the produced nuclei (a few MeV) is low compared to the temperature of the hadronizing system at the kinetic freeze-out ($T_{kin} \sim 10^2$ MeV) in which they are immersed. Two phenomenological models are currently available to describe the measured production yields, providing very different interpretations for the experimental observations. In the Statistical Hadronization Model (SHM) [1], hadrons are produced by a thermally and chemically equilibrated source and their abundances are fixed at the chemical freeze-out. This model provides a good description of the measured hadron yields in central A–A collisions [2]. However, the mechanism of hadron production and the propagation of loosely-bound states through the hadron gas phase are not addressed by this model. On the other hand, the production of light (anti)nuclei can be explained via the coalescence of protons and neutrons that are close by in phase space at the kinetic freeze-out and match the spin, thus forming a nucleus [3]. The key parameter of the coalescence models is the coalescence parameter, which is related to the production probability of the nucleus via this process and can be calculated from the overlap of the nucleus wave function and the phase space distribution of the constituents via the Wigner formalism [4].

In this contribution, recent results obtained by the ALICE Collaboration in Pb–Pb collisions are presented and compared to the expectations of models.

## 2. Ratio of nucleus and proton integrated yields

In order to extract the light (anti)nucleus integrated yields, transverse momentum ($p_T$) spectra are measured and extrapolated in the unmeasured low- and high- $p_T$ regions by means of a fit with a Blast Wave (BW) function [5].

The ratio between the measured yields of nuclei and that of protons is sensitive to the light nuclei production mechanism. In Fig. 1 the yield ratio to protons for deuterons (left panel), $^3$H and $^3$He (right panel) as a function of $< \mathrm{d}N_\mathrm{ch}/\mathrm{d}\eta_\mathrm{lab} >$ measured in pp, p–Pb and Pb–Pb collisions [6 - 9] is shown and compared to the expectations of the models. A smooth increase of this ratio with the system size is observed, reaching constant values in Pb–Pb collisions. The two ratios show a similar trend with $< \mathrm{d}N_\mathrm{ch}/\mathrm{d}\eta_\mathrm{lab} >$, however the increase from pp to Pb–Pb is about a factor 3 larger for $^3$He/p than for d/p. The observed evolution of the d/p ratio is well described by the coalescence approach because of the increasing phase space in Pb–Pb. For high charged-particle multiplicity densities, the coalescence calculations and the canonical statistical model (CSM) expectations are close and both describe the behaviour of the data, within the current uncertainties. On the other hand, the models struggle to describe the ratio to protons for nuclei with $A = 3$, as it is clear in the right panel of Fig. 1.

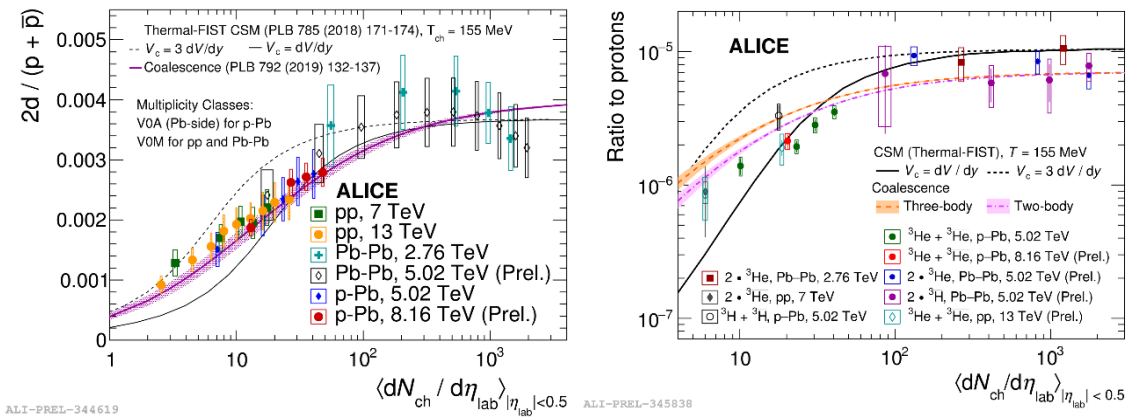

**Figure 1**: Deuteron (left panel), (anti)$^3$He and (anti) $^3$H (right panel) over proton ratios as a function of the charged-particle multiplicity in different collision systems and energies. Statistical uncertainties are represented as vertical lines whereas boxes represent the systematic ones. The results are compared to the expectations of SHM and Coalescence models.

### 3. Coalescence parameters

The coalescence parameter $B_A$ is given by the ratio between the invariant yield of the nucleus with mass number $A$ and that of protons, being $p_\mathrm{T}^\mathrm{p} = p_\mathrm{T}^A /A$, calculated as follows:

$$B_A = \left( \frac{1}{2\pi p_\mathrm{T}^A} \left( \frac{d^2 N}{dy dp_\mathrm{T}} \right)_A \right) \Big/ \left( \frac{1}{2\pi p_\mathrm{T}^\mathrm{p}} \left( \frac{d^2 N}{dy dp_\mathrm{T}} \right)_p \right)^A \qquad (1)$$

In Fig. 2 the coalescence parameters $B_2$ and $B_3$ as a function of the charged-particle multiplicity are shown, measured respectively for deuterons and tritons at fixed values of $p_\mathrm{T}/A$. The measurements show a smooth transition from low charged-particle multiplicity densities, which refer to a small system size, to larger ones. The decreasing trend of $B_A$ with increasing multiplicity suggests that the production mechanism in small systems evolves continuously as the one in larger systems and that a single mechanism sensitive to the system size could be responsible for nuclei production.

The theoretical calculations from coalescence [4] and the hydrodynamic-inspired Blast Wave model [10] qualitatively agree with the trend observed in data.

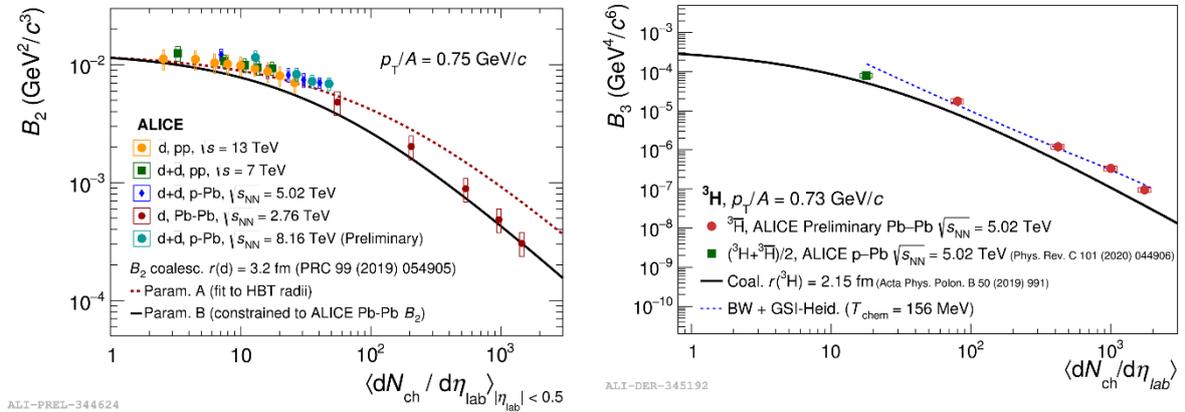

**Figure 2**: *Left.* $B_2$ as a function of $< \mathrm{d}N_\mathrm{ch}/\mathrm{d}\eta_\mathrm{lab} >$ for a fixed value of $p_\mathrm{T}/A = 0.75$ GeV/$c$. The experimental results are compared to the coalescence calculations from [4] using two different parametrizations for the system size as a function of multiplicity. *Right.* $B_3$ calculated for tritons as a function of the mean charged-particle multiplicity density for a fixed value of $p_\mathrm{T}/A = 0.73$ GeV/$c$. Results are compared to the coalescence calculations from [4] and to SHM coupled to hydrodynamic-inspired BW model [10].

## 4. Elliptic and triangular flow

Key observables to study the production mechanism of light (anti)nuclei are the elliptic and triangular flow, i.e. the second $(v_2)$ and the third $(v_3)$ harmonic, respectively, of the Fourier decomposition of their azimuthal production distribution with respect to a collision symmetry plane. The latter is defined by the impact parameter of the incoming nuclei and the beam direction. The flow of identified hadrons is often described using the hydrodynamic-inspired Blast Wave model [10], which assumes that the system produced in heavy-ion collisions is locally thermalized and expands collectively with a common velocity field. The system undergoes a kinetic freeze-out at the temperature $T_{kin}$ and is characterized by a common transverse radial flow velocity $(\beta)$ at the freeze-out surface. Elliptic flow can also be calculated in the simple coalescence approach, in which the deuteron $v_2$ is obtained from that of protons assuming that the flow coefficients are related as follows: $v_{2,d}(p_T) = [\ 2\ v_{2,p}\ (p_T/2)]/[\ 1 + 2\ v^2_{2,p}\ (p_T/2)]$.

The elliptic flow of (anti)$^3$He has been measured by ALICE in Pb–Pb collisions at $\sqrt{s_{NN}} = 5.02$ TeV as a function of the transverse momentum, for different centrality classes [11]. The $v_2$ is shown in Fig. 3 and compared to the expectations of models. As it is clear from the figure, the data lie between the expectations of models with a dependence on the centrality class: the simple coalescence approach describes the most peripheral collisions better, whereas most central ones are better described by the BW model. The same behaviour is also observed for (anti)d elliptic flow [12]. Therefore, the simple models considered here bracket a region where the light nuclei $v_2$ is located and describe reasonably the data in different multiplicity regimes, indicating that none of these two models is able to describe the production measurement from low to high multiplicity.

More sophisticated coalescence calculations, which combine a hydrodynamical simulation (iEBE-VISHNU) with later coalescence of protons and neutrons [13], provide a good description of the measurement in the centrality ranges below 40% not only for deuteron $v_2$ and $v_3$, as can be seen in Figure 4, but also for the elliptic flow of $^3$He [11]. Nevertheless, currently there are no predictions for coalescence results for more peripheral events.

## 5. Conclusions

The coalescence approach describes the experimental results concerning the ratio of the integrated yields of nuclei and protons as well as the coalescence parameter $B_A$ as a function of the charged-particle multiplicity density at midrapidity. For high charged-particle multiplicity densities, the coalescence approach and the CSM both succeed in the description of the d/p ratio, whereas models struggle to describe the ratio to protons for nuclei with $A = 3$.

Simple coalescence and hydrodynamic-inspired Blast Wave models are used to compare the results of elliptic and triangular flows of deuteron and $^3$He measured in Pb–Pb collisions at $\sqrt{s_{NN}} = 5.02$ TeV. These models bracket a region where the flow coefficients of light nuclei are located and describe reasonably the data in different multiplicity regimes. Additionally, experimental results are compared to more advanced hydrodynamical simulations (iEBE-VISHNU) with later coalescence, in the centrality region below 40%, where such predictions are available.

These results indicate that additional efforts both on experimental and theoretical side are needed to fully understand the production of light (anti)nuclei.

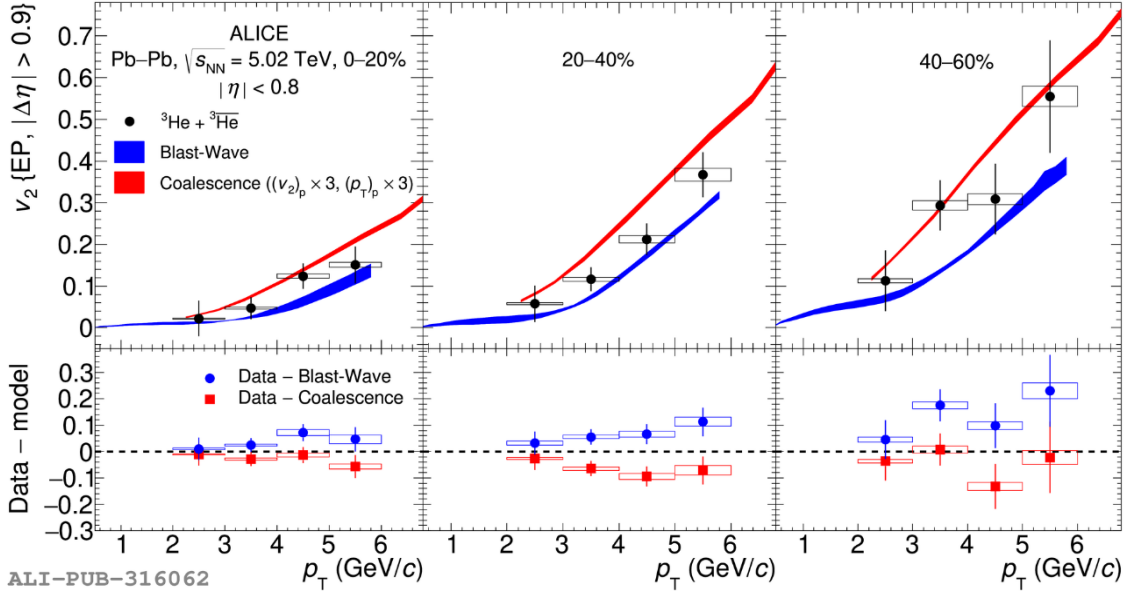

**Figure 3**: Elliptic flow of $^3$He as a function of $p_T$, for the centrality classes 0-20%, 20-40% and 40-60%. The experimental results are compared to the simple coalescence calculations as well as to the simplified hydrodynamic-inspired Blast Wave model [10]. The data-to-model ratios are shown in the bottom panels. Vertical bars and boxes represent the statistical and systematic uncertainties, respectively.

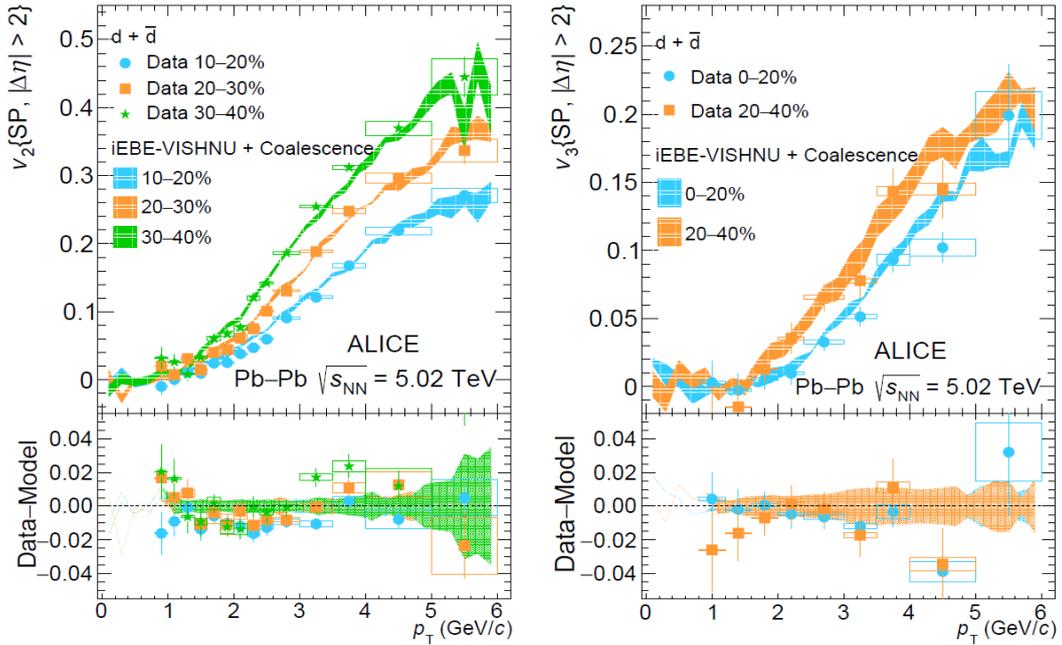

**Figure 4**: Elliptic (left) and triangular (right) flow of deuterons compared to the predictions iEBE-VISHNU hybrid model with AMPT initial conditions [13]. The predictions are shown as bands whose widths represent the statistical uncertainties associated with the model. The data-to-model ratios are shown in the lower panels. Vertical bars and boxes represent the statistical and systematic uncertainties, respectively.